\documentclass[preprint,showpacs,preprintnumbers,12pt]{revtex4}

\usepackage{graphicx}
\begin{document}
\title{Tailoring the physical properties of 
 poly(3-Hexylthiophene) thin films using electro-spray deposition.}
\normalsize

\author{M. Ali$^1$, M. Abbas$^{1,2}$, S. K. Shah$^1$, E. Bontempi$^3$, P. Colombi$^4$, A. Di Cicco$^1$ and R. Gunnella $^1$}

\affiliation{
$^1$ Sez. Fisica UdR CNISM, Scuola di Scienze e Tecnologie,
 Universit\`a di Camerino, via Madonna delle Carceri, 
I-62032 Camerino (MC), Italy\\
$^2$ Linz Institute for Organic Solar Cells, 
No. 69, Altenberger Strase,4040, Linz, Austria\\
$^3$Chemistry for Technologies Laboratory,
University of Brescia, via Branze 38, 25123, Brescia, Italy\\
$^4$Centro Coating C.S.M.T. Gestione S.c.a.r.l.
Via Branze, 45 I-25123 Brescia }

\begin{abstract}
Structural and electronic properties of homogeneous poly(3-Hexylthiophene) 
(P3HT) films obtained from the electro-spray method
 were presented by means of grazing incidence x-ray
diffraction, atomic force microscopy, optical absorption,
 photoelectron spectroscopy and (photo)electrical conductivity. 
Different structural conformations were obtained starting from different solution
concentrations and flow rate conditions.
The electro-spray method was shown to effectively expand the control 
of the conformation and assembling of the polymers films, opening the way to the
possibility of tailoring film characteristics according to device
specifications.

\end{abstract}

\maketitle
\section{Introduction}

Very recently, bulk heterojunction organic solar cells 
 \cite{sariciftci} using conjugated polymers blended with other inorganic 
nanostructured materials  have achieved a particularly refined level 
of process versatility, 
cost-effectiveness and high efficiency \cite{shaheen,reyes}. 
 Furthermore this field has given  
the intriguing perception of a wide room  available for further progresses. 

Poly(3-Hexylthiophene) (P3HT) is an important  
photo-conductive polymer for photovoltaic applications. 
The reason is mainly because of its
high drift mobility. Furthermore flexible side alkyl chains
improve greatly the  solubility, promoting solution processibility and
preserving $\pi$ electronic structure  \cite{bao,sirringhaus}.
For this reason, many efforts to maximize the efficiency of organic 
solar devices started from this polymer and  
aimed, on one side, to
lower the semiconductor polymer band gap  \cite{sariciftci2,hoppe}, on the other
 side,
to improve crystallinity by thermal
 \cite{sariciftci3} or solvent annealing treatments  \cite{jo}.  
It is well known that crystallization
was able to considerably increase the photons harvesting and hole mobility 
in the polymer as demonstrated in standard drop cast or spin coated films. 
More recently, attempts 
to optimally match 
the nanometer-scale morphology resulting from novel growth techniques
 and the basic microscopic processes in the device \cite{ma} have been proposed. 

Advances in growth techniques of thin active films is a key 
issue to this progress for a twofold reason.
On one side, looking forward to future applications, mainly to 
flexible substrates and in general for large scale implementations,
 novel deposition procedures have been proposed
like inkjet \cite{inkjet} or doctor blading \cite{blading} , which 
are easily scalable to roll-to-roll production. In particular
 the spray coating \cite{spray,susanna,green,ishikawa,vak} 
is  broadly employed to 
reduce the fluid waste to minimal 
amounts, while the combination of shadow masking and the control in the growth allows
 easy patterning and composite multilayered films fabrication to further improve efficiency.
On the other side, the search for valid alternatives to the few 
active materials in use today, like the polythiophene-fullerene blends, suggests to try different
paths in the device fabrication and to implements new processes.

The present study was focussed on the electro-spray deposition (ESD) technique
 \cite{jaworek,swarbrick,dam,cascio,tang} in order to explore 
its potentiality in the growth of thin films with properties possibly at variance 
from those of the P3HT 
obtained from other well assessed techniques like spin-coating.

In particular in recent works implementations 
 of the electro-spray
technique to the fabrication of photovoltaic bulk heterojunction devices were showed
\cite{tang,park2,kim2}.

Intrinsic drawbacks of the ESD technique 
 like roughness and the thickness profile,
 should be carefully considered and discussed 
to comply with a successful device implementation. As recently discussed by Wong and 
coworkers \cite{wong} spray methods giving rise to highly 
rough films with lower charge carrier mobility, 
can be safely cured after growth by post-solvent vapor annealing.

In the past, the ESD technique was variously applied  to several kinds of
solutions of biomacromolecules and/or synthetic polymers under strong 
electric fields giving rise to nano-sized particles
and fibers,
which accumulated and adhered on a substrate in ambient
conditions \cite{jaworek}, as well as in ultra high vacuum \cite{swarbrick}.

This transformation was brought about by the application of a high
voltage to the capillary through which the liquid
flowed. An important advantage of the technique  \cite{tang},
was the control that can be achieved on the
particle size, ranging from the micron to the nano scales keeping an almost monodispersed 
distribution, affecting differently the final device properties.
Furthermore, the resulting particles isolated from the solvent were supposed to experience 
different aggregation mechanisms from those observed in presence of solvent; such mechanisms could 
possibly be exploited to improve the efficiency of devices, and more importantly 
to widen the choice of possible blend components. 

For this reason, in this study we presented results  from  a low temperature 
boiling point solvent like chloroform, that  will enhance the 
electro-spray peculiarity of reducing at the best the  interaction 
of solvent and deposited films. 
 
Another possible advantage of the present technique will be the contribution of the ESD technique to  
fabricate  devices with complex heterostuctures by using different 
materials and/or solvents, or differently aggregated phase of the same
 active material. 
Typical examples are the optimized device architectures of a solar cell  or  
of a field effect transistor, taking advantage of the anisotropy in mobility 
resulting from a different $\pi-\pi$ stacking arrangement.
In general, sensors and other devices would receive a sizeable impulse
from a supramolecular organization of the characteristic two-dimensional 
$\pi$ states system of conjugated polymers \cite{kim}.

In brief, the present study will regard the peculiar physical properties of  
as-grown thin films obtained from the ESD methods applied to 
P3HT-chloroform solution at various electric fields, 
solution concentrations and deposition rates, 
while postgrowth treatments or 
 solvent choice effects will be the subject of future studies.    

Microscopic investigations were performed by
atomic force microscopy (AFM) and grazing
incidence x-ray diffraction (GIXRD), respectively for the morphology and 
the structure of the crystalline part of the polymer structure. 
The optical and
electronic properties were investigated by means of optical absorption in the visible range and
ultraviolet photoelectron spectroscopy (UPS). Finally, we observed the 
resulting conductivity 
obtained for such films by I-V curves at room temperature (RT), 
including illumination conditions and its wavelength dependence by photoconductivity (PC).

\section{ Experimental section}

Thin  P3HT films with thickness ranging between 100 and 200 nm, as measured by 
AFM after scratching the samples, were obtained 
 from chloroform solution of commercial electronic grade regioregular ( $>95\%$)
powders (Sigma), average molecular weight in excess to 50kD 
without further purification.

\begin{figure}
\includegraphics[width=0.8\textwidth, angle=0]{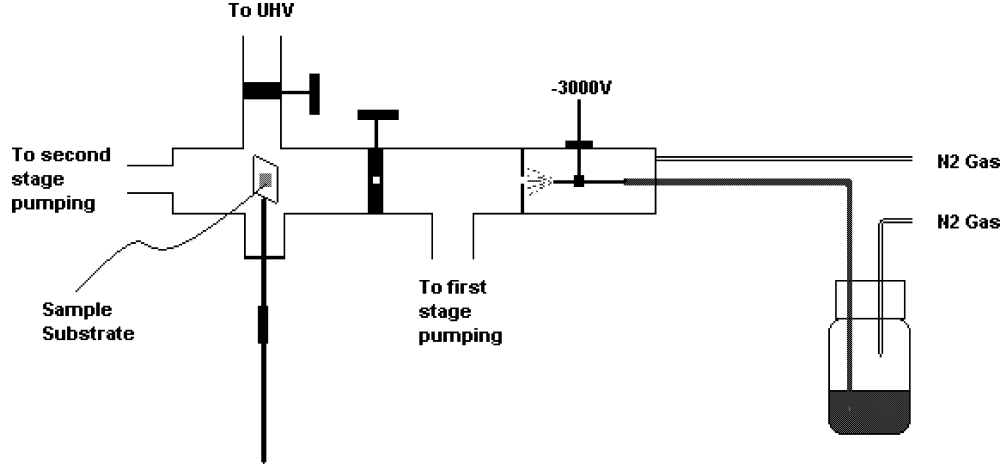}
\centering
 \caption{ Electro-spray system employed for the deposition of
 the P3HT films.} \label{gun}
\end{figure}

The homemade ESD apparatus is described in Fig. \ref{gun}. 
Stainless steel tip (inner diameter 0.1 mm) was modified before each use
 through electrochemical etching in the $H_{2}SO_{4}:H_{2}O, 1:1$ solution
to obtain a homogeneous electric field spatial distribution.  

The tip was centered to the pin-hole in the
grounded skimmer with high accuracy. Distance between tip and skimmer 
was 2-3 mm.  

When finely tuned $N_{2}$ gas over-pressurized
the solution container, a controlled flow of solution could reach
the spray tip, and electro-ionization occurred (so called Taylor
cone). Extracted from the grounded skimmer, the molecules were
driven to the substrate using a differentially pumped system (the
first stage at 1 torr and the second stage at $10^{-3}$ torr)  by a 
system of few millimeters diameter pin-holes one acting as a skimmer the other 
one before the sample holder. The sample holder allows heating during the 
growth or during post-growth annealing. The solution flow rate was measured by  
the time required to move a fixed amount of solution volume in the tube. 
It was found that the optimally applied positive voltage to the tip was 
 in the range of 1-3kV, depending on the present design of the reaction chamber 
(distance of tip from the grounded skimmer
and the pumping system).  Below such a range, the solvent was not removed
 efficiently from the solution and the flux inside the tip tended to be blocked.
Furthermore, concentrated solution at low rate led to discontinuous
flow with dark spots observed on the samples indicating polymer agglomeration.
 In the present study 
this effect was observed at 2 kV voltage and 1.3 $\mu l/s$ rate in the case of 0.5
mg /ml solution. On the other hand, above a voltage of 3kV, carbonization effects in the films
 looking dark and dusty were identified by the help of photoemission.

 The films were deposited on several substrates, namely,  corning glass, 
Indium Tin Oxide
(ITO) and on PEDOT:PSS/ITO substrates at two different solution concentrations of 
0.1mg/ml and 0.5 mg/ml, hereafter referred as    
type I and type II samples respectively.  
  After the
preparation, thin films obtained from a fresh solution resulted to
be stable if properly preserved in dark and controlled
ambient temperature. The morphology of the P3HT films was analyzed
by AFM both in contact and non-contact mode. AFM quantitative morphological
analysis was performed using Gwyddion software.
Absorption  spectra were obtained by using a lock-in amplifier and
an optical chopper. The working frequency chosen was 30Hz.
Absorption measurements were carried out by measuring the
normalized transmitted light intensity using a Silicon PIN
detector.

Grazing incidence X-ray diffraction (GIXRD) was collected by means
of a D8 Advance Bruker diffractometer
 keeping fixed the incidence angle at  $0.2^{\circ}$ during the detector 
scanning in the plane of incidence.

UPS spectra were acquired at pressure of $\sim
10^{-9}$ Torr using a He discharge lamp providing HeI (21.2 eV)
and HeII (40.8 eV) lines impinging at $45^{\circ}$  with respect to the
detection direction of a hemispherical electron analyzer
VG-CLAM4 or by rotating the polar angle of a three axes sample manipulator.  
HeII line allowed to collect spectra with reduced influence of the secondary electron emission. 

The photoconductive signal was measured over the load resistance of 
$20M\Omega$ under an electric field of $5.3 \cdot10^{3}V/cm$.
The incoming light power was derived from the standard photoresponse 
of the silicon detector
to normalize the photo-response spectra of the samples.

Finally, the electrical characterization  was
carried out in dark and illumination conditions (ambient light)
  by a Keithley 617 electrometer,
in the V/I mode, using aluminium stripes evaporated onto 7059
Corning glass substrates after film deposition.  Alternatively for sandwiched structures 
ITO covered glass was spin coated by PEDOT:PSS and used as substrate. Thin gold
wires were put in contact with the metal stripes by using silver paint,
 dried in dark. The sample thickness was 100nm while electrode cross-section 
was 2 mm with separation gap of 150 $\mu m$.

\section{Results}

\subsection{Microscopy}

\begin{figure}
\centering
\includegraphics[width=1.0\textwidth,angle=0]{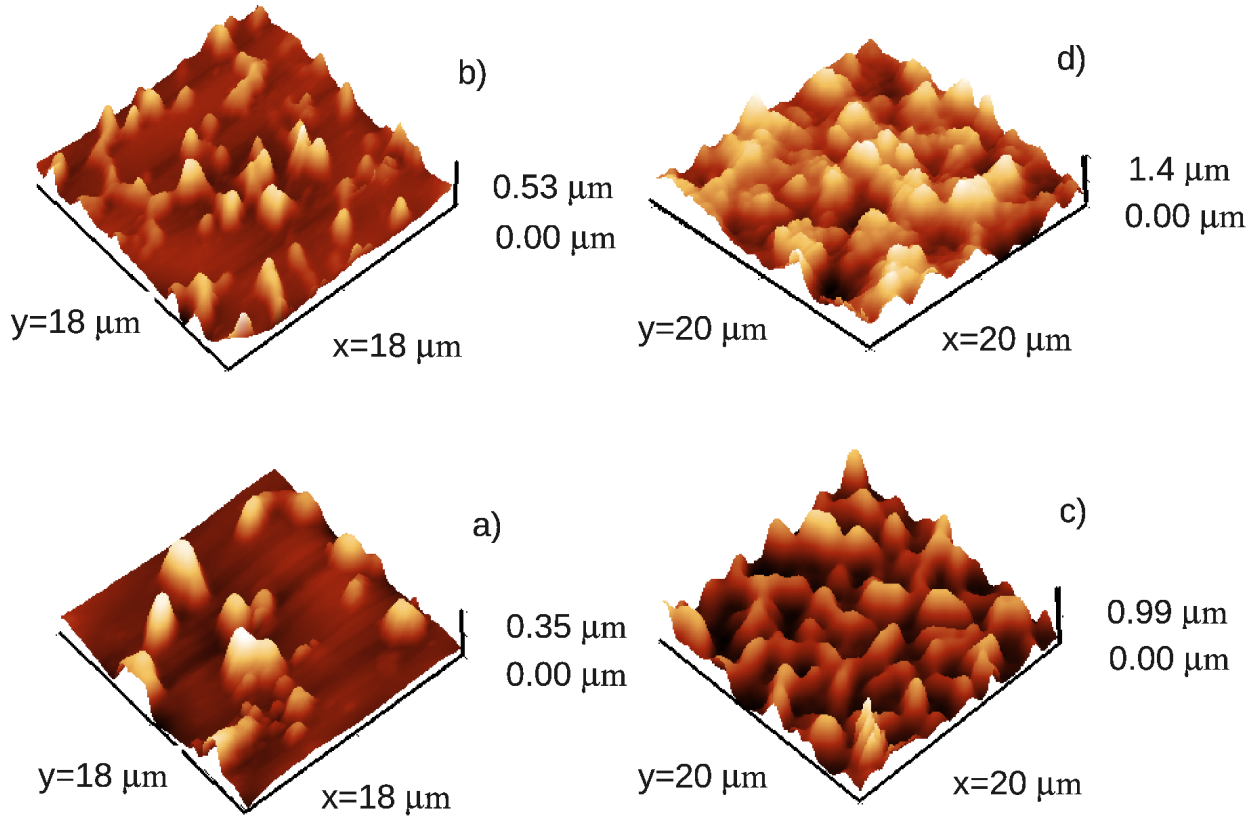}
\caption{ AFM images of electro-spray (2kV) deposited P3HT films, 
from 3 ml solution at different flow rates are reported.
 For a concentration of 0.1 mg/ml from a) to b) the solution flow speed 
 is 1.3 and 4.0 $\mu l /s$ respectively.
 For 0.5 mg/ml solution pictures c) and d) are from samples grown with
  flow rates of 2.0 and 4.0 $\mu l$ respectively.} \label{afm}
\end{figure}

AFM images  of type I samples obtained from a constant amount of solution 
(3 ml)  at the  flow rate  of  1.3 and 4.0 $\mu l/s$  and for type II at the rate 
of  2.0 ( the minimum achievable with good quality
at this density ) and 4.0 $\mu l/s$,  were 
reported in Fig. \ref{afm}a)-b) and in Fig. \ref{afm} c) and d) respectively. 

Though of dissimilar thicknesses the type I and type II films 
of Fig.\ref{afm}, showed peculiar differences 
 and a continuous evolution at the increase of flow rate and concentration
 from a high corrugation and anisotropic morphology 
( Fig.\ref{afm}-a) ) to a more isotropic growth ( Fig.\ref{afm}-d) ).

\begin{figure}
\centering
\includegraphics[width=1.0\textwidth,angle=0]{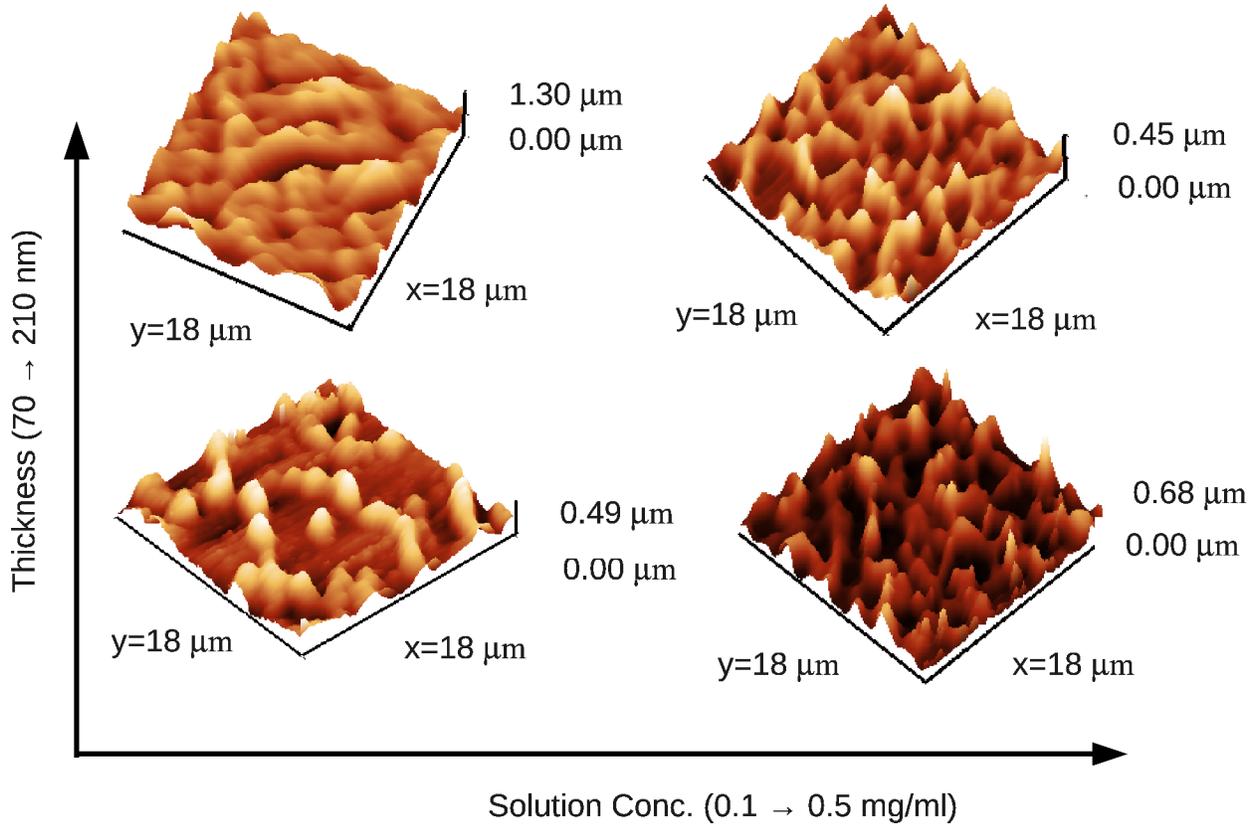}
\caption{AFM pictures showing the morphology dependence at different 
conditions of films grown  with a rate of 2.0$\mu l/s$ 
at solution concentration of 0.1 (left images) and  0.5 mg/ml ( right images)  
and sample thickness of about  70 ( lower images ) and 200 nm (upper images).} 
\label{afm2}
\end{figure}

Such an observation was not related to thickness difference of the samples
 as it was demonstrated by
the pictures reported in Fig. \ref{afm2} for 
flow rate of 2.0$\mu l/s$ where the combined effects of solution concentration
 and sample thickness of films grown were displayed. 
In fact  while type II samples showed a rough isotropic morphology, in type I 
samples a sizeable roughness was rather due to a less compact and anisotropic 
arrangement of islands.

\subsection{X-ray diffraction}

\begin{figure}
\centering
\includegraphics[width=0.6\textwidth,angle=0]{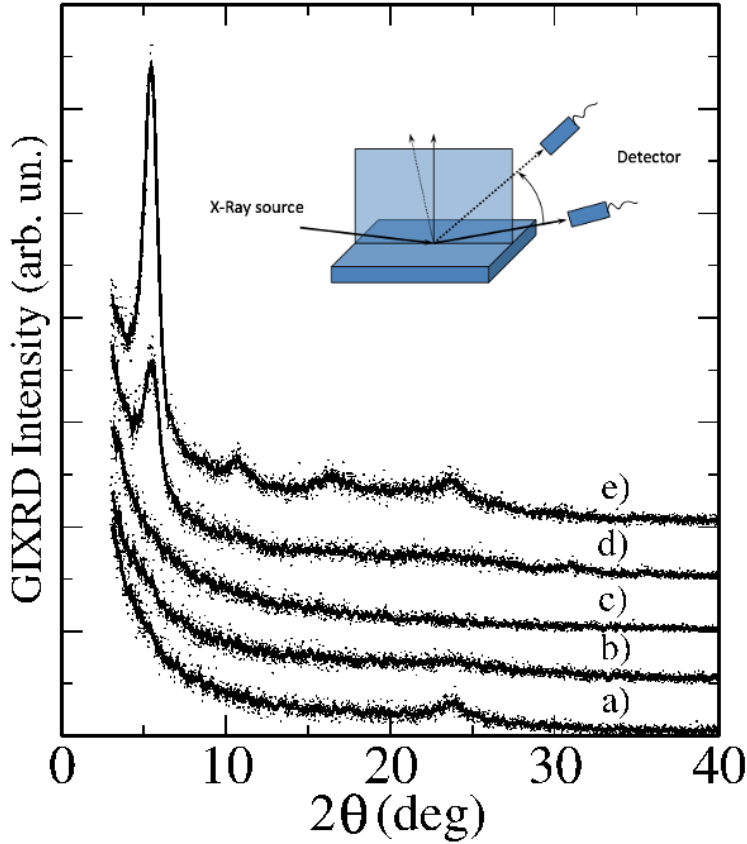}
\caption{ Grazing Incidence XRD from P3HT films: from  (a) to (c)  
the diffraction patterns of  0.1 mg/ml solution at the
 speed of  1.3, 2.0 and 4.0$\mu l/ s$ respectively. Curves (d)-(e) represent the patterns of  
 0.5 mg/ml solution samples grown at 2.0 and 4.0 $\mu l/s $ flow speed respectively. In the 
in-set is reported the sketch of the grazing geometry. }
\label{gixrd}
\end{figure}

For the structural determination of the P3HT films we
employed GIXRD, an
extremely suitable tool to determine the surface structure in those
systems whose physical properties of interest were confined at the
interface region or where modifications took place. To this
aim, a Cu K-alpha source and incidence angle of about $0.2^{\circ}$
  were used, while the detector
was scanned in the plane of incidence ( see in-set in Fig. \ref{gixrd}).
 Below an angle of $1^{\circ}$ of 
incidence a drastic reduction of the substrate contribution was
found in the patterns.  In fact, as for an 
incidence angle of about $0.12^{\circ}$, the X-ray penetrates theoretically only 10 nm
 in the film, and at $0.18^{\circ}$ about 60-70 nm, we  suppose that,
with the limitations of sizeable roughness observed ,  at the present incidence
conditions we reasonably increased the sensitivity to the film
 whose thickness was ranging between 150 and 200 nm. 

In the
symmetric $\theta/2\theta$ configuration, the scattering vector is
always parallel to the surface normal, and XRD is sensitive to crystallographic
 planes
parallel to the sample surface. In the present grazing incidence
configuration, neither scattering vector was  parallel to the
surface normal, nor its orientation remained constant during
measurement, then contribution was made by lattice planes that
made a non-zero angle with the substrate. In this manner, applying a very low 
angle of incidence  the
technique was more sensitive to crystallites with scattering vector
whose angle was lower than $90^{\circ}$ ( direction of the surface normal). 
GIXRD measurements were shown in Fig.\ref{gixrd}.
Within the unit cell model proposed by Prosa and coworkers \cite{prosa},
 P3HT crystallizes 
in a monoclinic unit cell, forming lamellae or planes of backbones perpendicular to
 thiophene rings and to the a-axis 
 ( a= 16.8 \AA, the interdistance of two lamellae or layers).
 Other lattice parameters are
  b= 7.6 \AA ( twice the stacking distance between two  thiophene rings),  c= 7.8 \AA ( the 
 distance
between two thiophene rings in the backbone) and $\gamma = 90^{\circ}$.
The thiophene ring plane is often denoted having either a "flat-on" orientation 
when the main portion of the crystalline component of the 
film has the (020) reflection oriented parallel to the sample normal or (h00) at $90^{\circ}$
to the sample normal. Alternatively an "edge-on" orientation of the crystalline portion of the film 
is invoked when the (020) reflection is at $90^{\circ}$
to the sample normal. In the latter case, the alkyl chains are sticking out of the 
sample surface.

The main diffraction features
recorded in our case, performing GIXRD at $0.2^{\circ}$ of incidence,
were the (h00)-reflection peaks at about $ 2\theta = 5^{\circ} $ and $10^{\circ}$,
corresponding to the inter-layer alkyl chains in the lamellar
structure ( a=16.8\AA\ ,
  q=0.37  \AA $^{-1}$  )  . 
A second feature was labelled (020) at $2\theta = 23.4 ^{\circ}$, corresponding to
2q = 4$\pi$ /b = 1.65\AA $^{-1}$  with  b=7.6 \AA  , due to intra-lamella distance 
 of the $\pi-\pi$ stacking [14].

In order to increase the portion of the reciprocal space 
investigated we performed conventional XRD measurements with an area detector.
XRD patterns ( not shown) reported only the peak at $5^{\circ}$, 
while that one at $23^{\circ}$ was
 missing maybe because of the sizeable roughness on the surface. Another possibility 
would be the reduced sample thickness. In any case the results 
pointed unequivocally towards the presence of an oriented crystal structure.   

In Fig. \ref{gixrd}, from bottom to top after suitable vertical 
shift of the plots, GIXRD patterns 
of type I samples at increasing flow speed of $1.3 \mu l/s$, 2.0 $\mu l/s$ and 4.0 $\mu 
l/s$ respectively (Fig.\ref{gixrd} from a) to c)) and of type II samples at   
 2.0$\mu l/s$  and 4.0 $\mu l/s$ flow speed  (Fig.\ref{gixrd} d) and e)) were reported . A clear
diffraction intensity ( the (020) peak) was observed   at $23^{\circ}$ during
low solution concentration/ low speed growth. The (h00) peak was missing because 
the scattering vector 
of the (100) reflection was perpendicular to the substrate.
Using a more concentrated solution, a clear indication of a planar thiophene ring 
was obtained. In this latter case,
the diffraction of (h00) orders was  observed due to parallel arrangement of molecular 
layers spaced by alkyl chains, but the peak at about $23^{\circ}$ 
was also observed at higher flow rate (Fig.\ref{gixrd}-e), maybe due
to the increase in misorientation.

\subsection{Optical absorption}

A vast literature is available on the optical absorption spectra
of P3HT. Absorption of solid state films shows remarkable red
shift with respect to that of the polymer in solution
\cite{patil}. When the solid film is formed from high molecular weight
regioregular polymer, molecular chains tend to develop from the
coil-like structure in solution \cite{neher} into more planar,
rigid rod-like morphology. Such a morphological change, together
with the extended conjugation length in the polymer backbone,
induces an extra delocalization of the $\pi$ states, and
consequent decrease of the excitonic energy. Another effect is the number of rotations (tilts) 
of backbone planes which stabilize the crystal structure \cite{northrup}. The main 
effects of this rotation were to decrease the bandwidth along the stacking direction 
by 25\% and a slight increase of the band-gap (0.1-0.5 eV)\cite{northrup}.

\begin{figure}
\centering
\includegraphics[width=0.8\textwidth,angle=0]{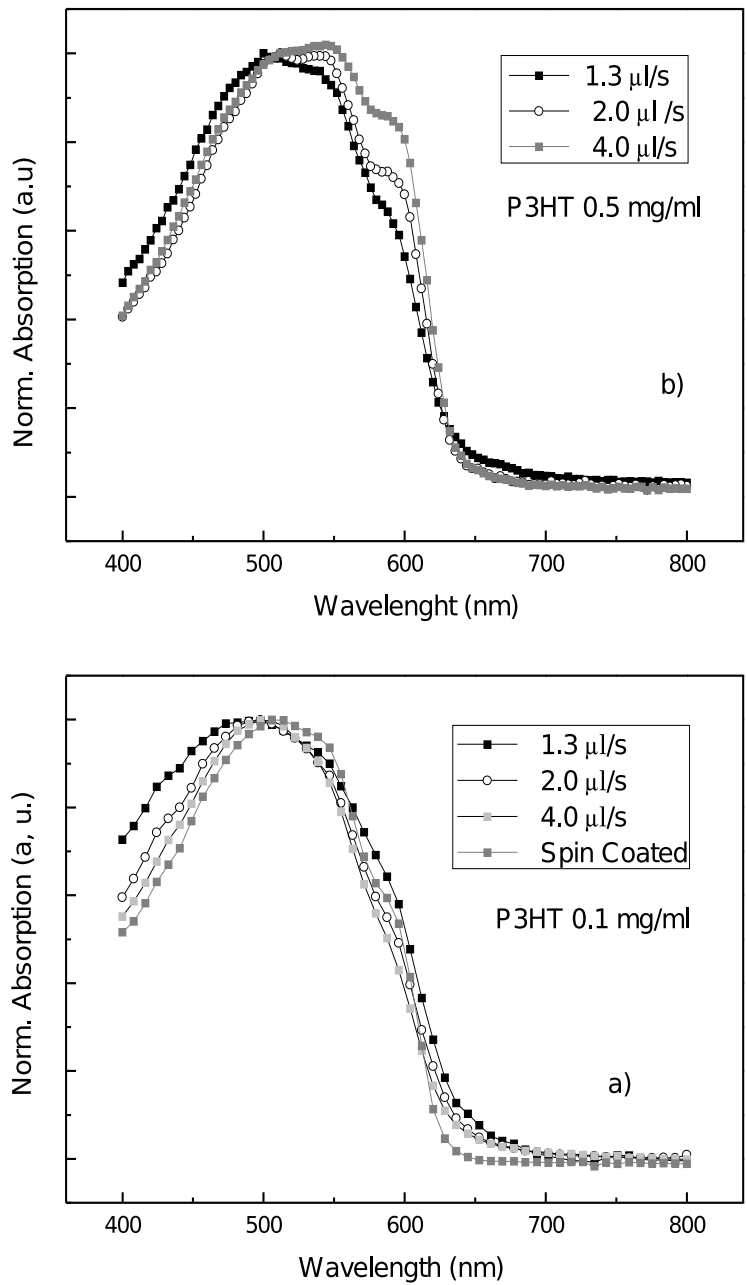}
\caption{ Absoption spectra of samples grown at different concentrations of the solution: 
 Panel a):  less concentrated 
solution (0.1 mg/ml) and speed of 1.3$\mu l/s$ , 2.0$\mu l/s$,  and
4.0 $\mu l/s$. Typical spectrum of a spin coated sample is also reported.
Panel b):  from  the 0.5 mg/ml solution samples grown at 1.3,  2.0 and 4.0 $\mu l/s $. 
  }
\label{abs}
\end{figure}

The absorption spectra collected on the present ESD films showed three main features: a) a first
shoulder structure at about 596 nm (2.08 eV); b) a second one at
about 549 nm (2.26 eV); c) a broad feature centered around 515 nm
(2.41 eV), which was the maximum of the spectrum used for the
normalization. In particular, the latter two structures were due to
the $\pi$ to $\pi^{*}$ intra-chain transitions of the $\pi$
conjugated polymer backbone. On the contrary, the low energy state
at 2.08 eV  (a) has a mixed contribution from inter-chain 
$\pi$ to $\pi^{*}$ \cite{brown}
transitions and from strong $\pi$-conjugation due to planarized 
 sections \cite{mccullough} along the main chain.

 As it can be observed in
Fig.\ref{abs} panel a), very similar absorption spectra were
obtained regardless the growth speed for type I samples.
 In the panel a)  was reported also the typical 
absorption spectra obtained for a spin-coated film \cite{abbas}, showing a comparable 
absorption spectrum.

At variance with these characteristics were the spectra
 from type II samples. 
In these cases spectra were reported in Fig.
\ref{abs}, panel b). In particular, the vibronic structure at 596 nm, 
and the feature due to 
conjugation length at 549 nm were enhanced with respect to type I samples. 
 This observation suggested a stronger planarization
and better $\pi $- $\pi$ interaction when a fast aggregation was occurring 
$(4.0 \mu l/s)$ from a more concentrated solution. 

The spectral width of Fig.\ref{abs}-b) increased indicating a sizeable amount of 
torsion and inhomogeneous conjugation in larger solution concentration and flow rate samples
\cite{kobashi}.

In the work of Park et al. \cite{park} the
ordering transition ( with substantially the same
effects seen here in the GIXRD and optical absorption) was
associated with thermal annealing above the melting point of the
film. 
We exclude that those 
effects could derive from temperature annealing experienced during the
 spray mainly because higher flow rates means a less exposure to the
high electric field, and because recent experiments 
of annealed P3HT above the glass transition temperature, reported only an increment in the 
absorption with a preserved shape of the absorption spectrum \cite{huang}.

\subsection{Photoelectron spectroscopy}

Photoelectron data were collected on films grown on conductive ITO substrates,
in order to avoid charging and smearing effects in the spectra.

In general, an oxidation free P3HT film \cite{shimo},
shows  well resolved photoemission peaks \cite{abbas} between 0.5 and  1.0  eV in binding energy 
(BE) attributed to the delocalized all $C2p_{z}$-$\pi$ states, while features 
localized between 1.0 and 2.5 eV BE have major contribution from $C2p_{z}$
and from $S3p_{z}$ atomic orbitals. In addition, the peak located around 3.8 eV
 includes localized $\pi$ states with $S3p_{z}$ and weak contributions from atomic
 $C2p_{z}$ orbitals \cite{hao}.

When looking at deeper energy levels, they are
dominated by the alkyl chain $\sigma$ states ( not
affected by the polymer backbone structural changes) between 6 eV and 10 eV BE. In particular,
 the contributions from the aromatic backbone $\sigma$ states are localized at
the lower energy side and the localized $\pi$ states in the higher energy
region. Finally, at 13 eV BE mainly electronic states
 with C2s character \cite{salaneck,chua} are
observed.

\begin{figure}
\centering
\includegraphics[width=1.0\textwidth,angle=0]{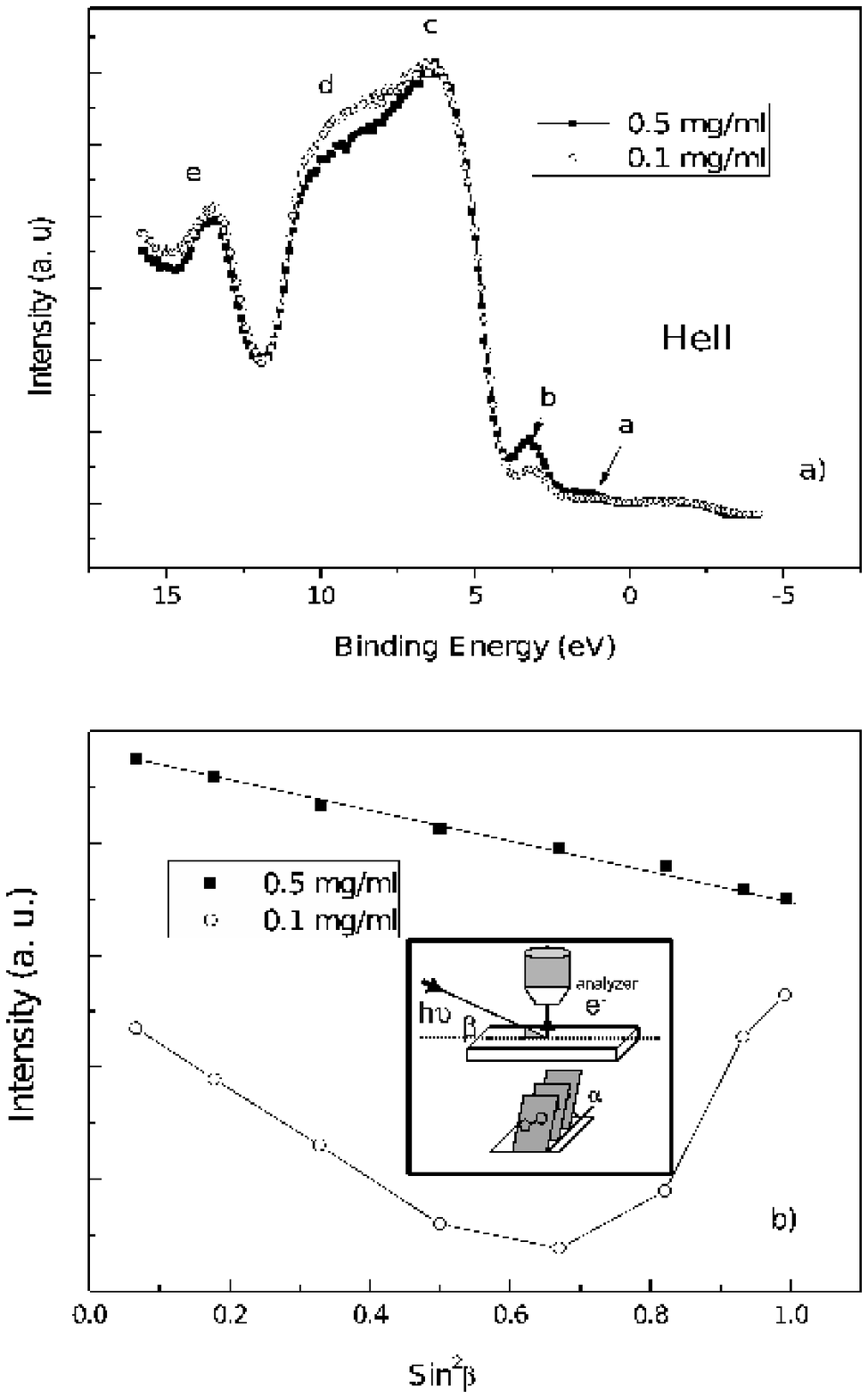}
\caption{ a) Normal photoemission (HeII)  data of the sample grown at low  
concentration/low speed (type I samples depicted by circles) and high  concentration/high speed
 conditions ( type II, squares) at 2kV voltage. The angle between the light and the detection direction 
( normal emission ) is $45^{\circ}$. b) Normalized intensity behavior of the localized $\pi$ states 
at 3.8 eV BE, as a function of the grazing angle $\beta$ of photon incidence for the type I ( open
symbols) and type II samples ( full symbols). In the inset is shown the pictorial sketch of 
the photoemission experiment during $\beta$ scans.  }
\label{ups}

\end{figure}
Furthermore, photoelectron spectra might give interesting hints about the conformation of the film
and the orientation of the molecular orbitals when related to a variation in 
spectral intensity in the region of localized states.
\cite{abbas}.

We reported in Fig.
\ref{ups} panel a)  well resolved HeII UPS spectra taken at
$45^{\circ}$ of polar angle, showing a higher intensity of $\sigma
$ states at 9 eV BE ( peak labelled "d") for type I samples (plotted by circles), 
while $\pi$ states at 3.8 eV ( peak "b") 
were enhanced in the case of type II samples ( plotted by squares). Following a recent study
 by Tao and coworkers
 \cite{tao}, such a behavior might be a prove that in type I samples, alkyl 
chains shielded
the photoemission from the thiophene rings by protruding towards the vacuum,
 indicating an edge-on 
geometry of the P3HT backbone. On the other side, unshielded
photoemission from the flat and exposed  $\pi$ system occurred in the case of type II 
samples.

A second more quantitative evidence came from the angular study of the localized 
$\pi$ states at 3.8 eV BE as a function of the 
grazing angle of incidence ( $\beta$ ) of the He II photons, as reported in Fig.\ref{ups} panel b).

The geometry of the experiment was sketched in the inset of panel b) of  Fig.\ref{ups}
 showing respectively the photon incidence angle $\beta$ and
the declination $\alpha$ of the thiophene rings with 
respect to the horizontal plane.
By the angular dependence in UPS spectra,
 different behavior in the case of the two sets of samples was found.

Features from 0.5 to 3.75 eV which are due to delocalized and localized 
 Cpz and Spz states perpendicular to the polymer backbone were fitted with Voigt 
functions after the subtraction of the linear background and properly normalized to avoid 
thickness effects. 
The intensities of the areas at different incidence angle $ \beta$
were plotted against $(sin\beta)^{2}$.
A linear trend was observed in case of type II ESD films. 
The linear fit was used to extrapolate intensities at 0 and $90^{\circ}$ incidence, 
which were then used to calculate the figure-of-merit 
R = (I(90) - I(0))/(I(90) + I(0)) yielding a value of -0.15.
 This indicated that high concentration ESD film had a tendency to form 
a major face-on configuration, while a purely face-on structure gave a value of -1 \cite{gurau}.
 However, the linear trend was not observed in case of low concentration ESD films,
 with a minimum at $55^{\circ}$, an indication that the film structure 
was showing a mixed configuration without either face-on or edge-on dominant phases
 but quite far from having an isotropic configuration 
that would have led to a constant contribution. Similar  mixed 
population was found by X-ray absorption in a recent work by 
 DeLongchamp and coworkers \cite{delongchamp} and put in relation
 to the presence of tilted section of polymer chains \cite{northrup}.





\subsection{Transport studies and photoconductivity}

\begin{figure}
\centering
\includegraphics[width=0.8\textwidth,angle=0]{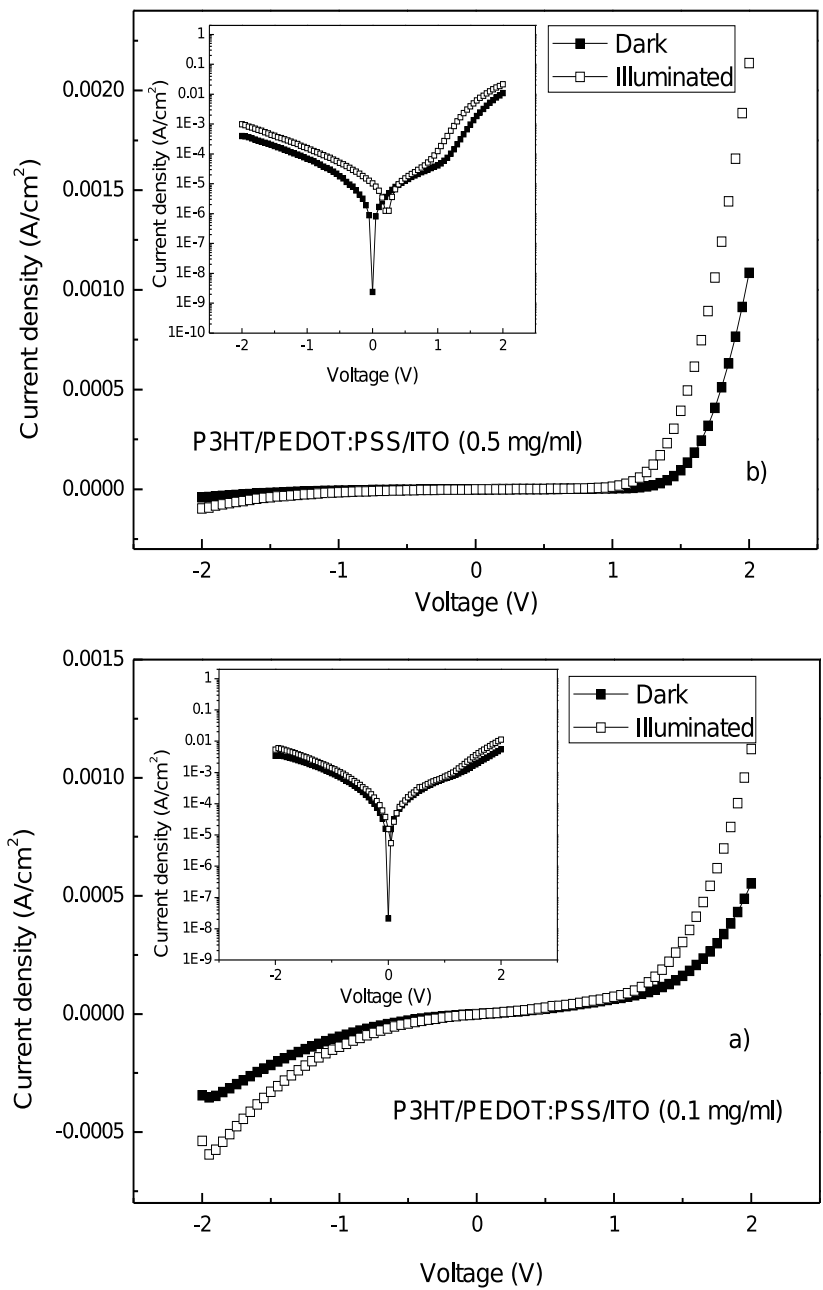}
\caption{ I-V curves of 100 nm  P3HT film sandwiched  between ITO/PEDOT-PSS 
and Al electrodes by  lower (a)  and higher concentration solution (b) 
respectively, in dark and ambient illumination conditions}. \label{diode}
\end{figure}

The transport properties were studied by planar and sandwiched geometries. 
The results for the latter case for samples grown by the two different solution concentrations
 were reported in Fig.\ref{diode} a) and b) respectively.

As it can be seen a clear diode-like behavior in the type II samples 
( Fig.\ref{diode} panel b) was found, while a less rectifying behavior was observed in 
Fig.\ref{diode} panel a), for the type I  samples. Probably the reason for the latter one
was the lower shunt resistance \cite{winder} 
because of the peculiar morphology and/or because of the high series resistance.

\begin{figure}
\centering
\includegraphics[width=0.8\textwidth,angle=0]{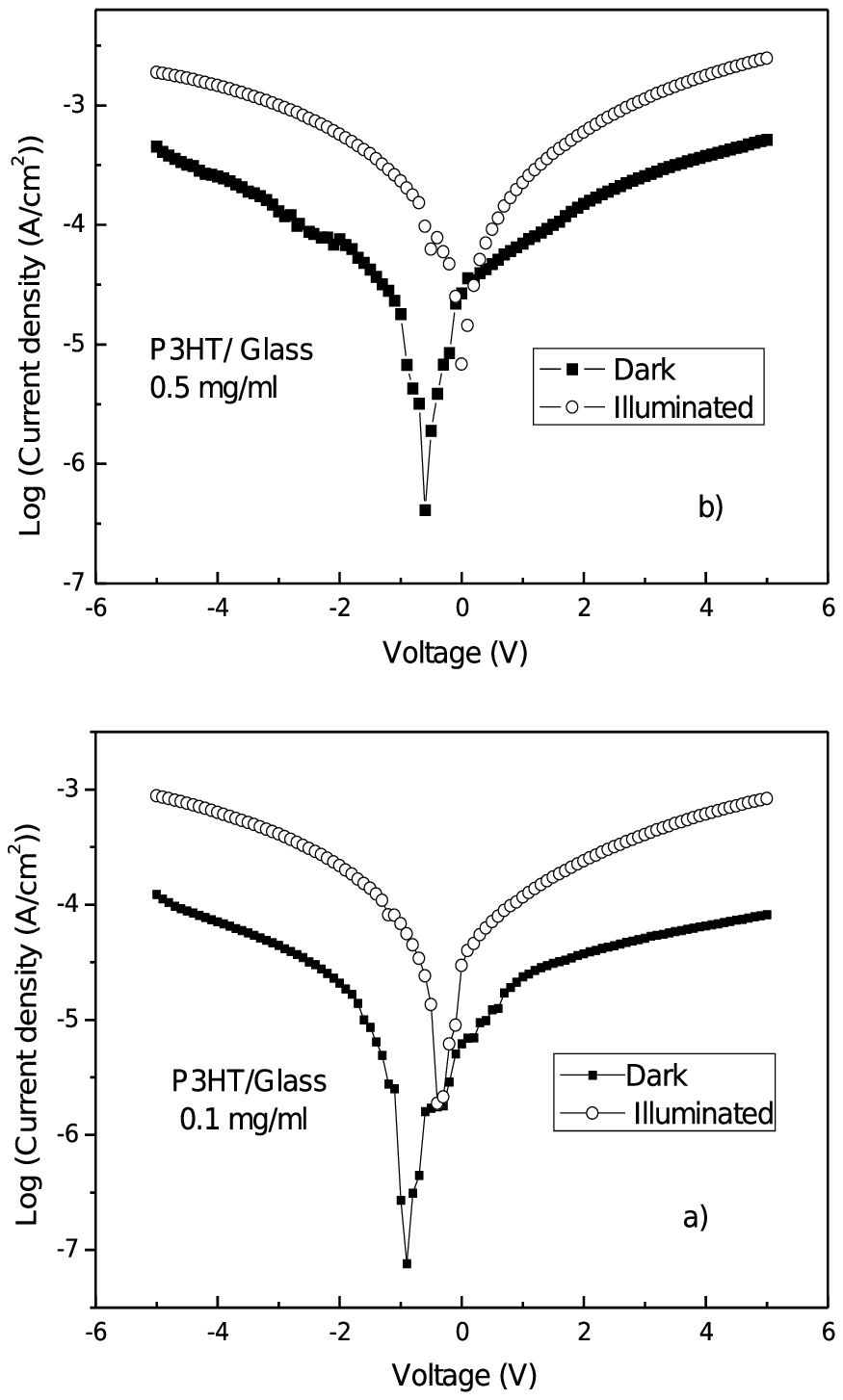}
\caption{I-V curves of planar structure of 100nm P3HT film grown between Al electrodes with 
150 $\mu m$ gap separation by lower and higher concentration solution  
respectively, in dark and ambient illumination conditions.} \label{mobility}                   
\end{figure}

A complementary information came from the planar conductivity in a different structure with in-plane 
electrodes.
At high electric fields of the order of $10^{5}V/cm$,
 the traps can be filled up completely 
and the space charge limited regime is dominant \cite{goh}. 
In the charge limited current approximation the mobility can  be described by the formula
$\mu= (8Jd^3)/(9\epsilon_0 \epsilon_r V^2) $ 
in terms of the dielectric constant ($\epsilon_{r}= 6.5$ ) 
 for the sample grown at low concentration and at high concentration ( flow speed 
$2 \mu l /s$).
By this analysis the higher concentration samples ( Fig. \ref{mobility} ) showed a mobility 
 three times larger than that obtained from  
 type I samples.
 Though such mobilities are still about two order of magnitudes below those ones 
 reported in literature from 
spin coated P3HT films \cite{giulianini}, a particular behavior of these samples  
by changing illumination conditions was observed. 
In fact, type I samples reported in Fig. \ref{mobility}-a) showed a strong photoeffect
with  increase in carriers under illumination close to 50\%. Such a 
sizeable response is more important than in type II samples.

\begin{figure}
\centering
\includegraphics[width=0.8\textwidth,angle=0]{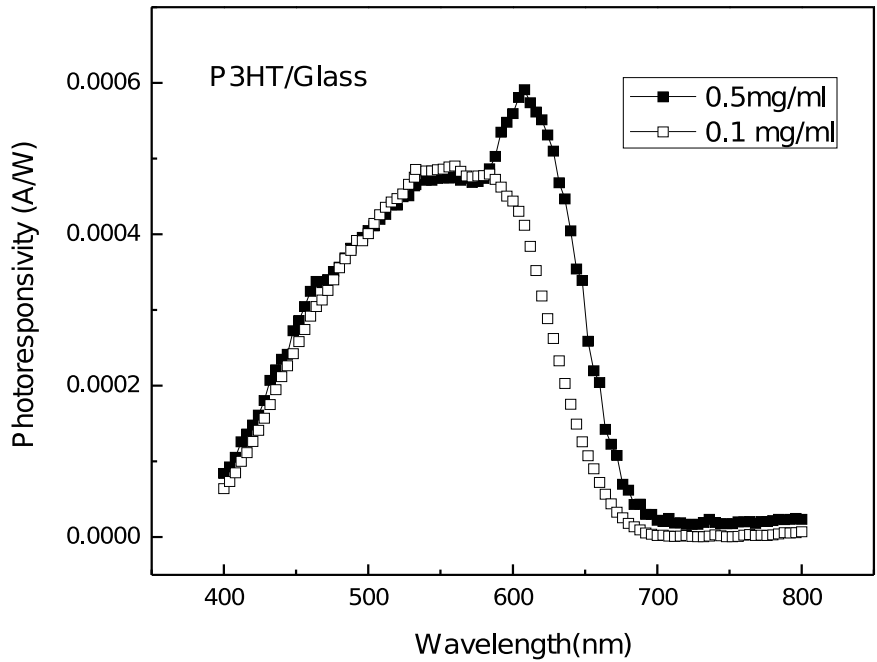}
\caption{  Spectral photoresponse after normalization to incident power for the higher and lower concentration 
solution samples grown on corning glass.  Al metallic contacts have
150 $\mu m$ separation gap.} \label{pc}
\end{figure}

The spectral photoconductivity normalized to the incident power  
confirmed  a strong absorption in the 600nm region ( see Fig.\ref{pc}) 
 in type II samples.

\section{Discussion and Conclusions}

The  structure of a P3HT films consists of a crystalline 
packaging of main conjugated polymer chains well oriented along the $\pi$ stacking direction. 
In such systems the formation 
of tilted or rotated structures has been found to reduce the total energy of about 0.4 eV/monomer 
\cite{northrup}, at the expense of the crystalline mobility 
by increasing the effective electron mass.     

 In fact, high in-plane mobility in P3HT films was reached in "edge-on" conformations 
of the polymer chains with thiophene rings perpendicular
with respect to the substrate \cite{sirringhaus}. Such a conductivity was found about three
orders of magnitude larger than the conductivity in films with
"flat-on" oriented chains.  The difference was given by the hopping through 
$\pi $ orbitals along the stacking direction oriented towards the conduction channel.
This was the condition
 obtained in the case of low molecular weight polymer films 
grown by  casting or in some cases by spin coating. Actually, if the regioregularity was
 reduced and/or the 
molecular weight increased, a flat-on growth was obtained with reduced 
crystallinity. Nevertheless  
 a general loss of crystallinity is not always detrimental to the
overall conductivity. In fact, a better connectivity is obtained by means 
of  enhanced conjugation length and the complexity of the structure overcomes 
the limitation posed by the 
grain boundaries \cite{kline} responsibles of  
the poor overall conductivity in low molecular weight P3HT fibrils.

In the present work we found that the electro-spray growth is regulated 
by both the solution concentration and the flow rate indicating that 
the electric field applied to each molecule is an important parameter for the growth,
able to tune the final characteristics of the film.

By the present study we found that the samples grown by a  
lower concentration solution and lower rates showed clear indications
of different conformation from samples obtained by more concentrated solutions 
 and larger flow rates.
These two cases represent the two extremes of the highest and lowest electric field perturbation 
that can be applied to the solute, within the range of operation of the present tool,
 depending largely on the  
mechanical design and the pumping system.   

  While at lower concentration solution we evidenced a non-planar 
orientation or edge-on ring conformation, at higher concentration the 
self-assembly of the molecules was pointing to a more planar orientation 
with a more isotropic continuous growth mode character. 

In fact, when the solution was less concentrated,
 self-assembly could be initiated by a
 stronger electric field 
experienced by the polymer with possible effects on the polymers alignement
and conjugation length. 

On the contrary when the flow speed was increased and the solution more concentrated
the effect of the ionization induced by the field was reduced accordingly, and 
 molecules interacted giving rise more 
rapidly to packaged films. 

Several observations supporting this view were presented in all the reported
experimental characterizations like the morphology dependence seen by AFM or the
orientation of the crystalline planes observed by GIXRD.

Similarly,  the peculiar behavior  of the 
optical  absorption with enhanced vibronic features \cite{kobashi},
in the case of high concentration solutions and high flow rate samples, or 
the angular anisotropy  of the localized valence band $\pi$ states are supporting this view.

In the conductivity measured in "sandwich" configuration, 
perpendicular transport through the $\pi$ stacking 
orientation was observed with a Schottky barrier characteristics 
in high concentration and flow rate solution samples.

In planar structures, conductivity measurements  
showed again the predominance of the higher solution 
concentration samples in spite of   
the unfavored packaging direction  of the  $\pi$ stacking, probably because of the longer conjugation 
length and of the reduced crystalline grain boundaries of a more isotropic 
and continuous film . Nevertheless  a more intense photoconductivity effect was observed
for low concentration solution films, indicating that of a more efficient photoabsorption 
was occurring because of the mutual orientation of 
the molecular orbitals and the incident electric field.

In conclusion, we showed a way to
vary  efficiently the properties of
self-assembling of a conductive polymer by 
finely tuning the speed of the growth and the solution concentration during electro-spray deposition 
of conductive polymers. In fact, we
observed a rapid transition from a prevalent edge-on building block of the
polymer to a flat-on configuration. This fact could in some way
remind the well known behavior of low/high regio-regular
assembling of spin coated or drop casting samples reported by
Sirringhaus and coworkers \cite{sirringhaus}, but in this case
such a trend was obtained from the same organic
material. Theoretical study of the atomistic process is now necessary to
understand how the details of the self-assembling are influenced by
the various parameters.

\newpage

\end{document}